\documentclass[aps]{revtex4}

\begin{document}

\title{Comment on ``Truncated Schwinger-Dyson Equations and Gauge Covariance in QED3''}

\author{Shang-Yung Wang}

\affiliation{Department of Physics, Tamkang University, Tamsui, Taipei 25137, Taiwan}

\date{17 March 2008}

\begin{abstract}
A comment on the paper ``Truncated Schwinger-Dyson Equations and Gauge Covariance in QED3'', Few-Body Syst.\ 41, 185 (2007) [hep-ph/0511291].
\end{abstract}
\maketitle

Bashir and Raya~\cite{Bashir:2005wt} have recently studied the gauge independence of dynamical chiral symmetry breaking in (2+1)-dimensional QED by implementing the nonperturbative Schwinger-Dyson (SD) equations in various truncation schemes. In two recent articles~\cite{Leung:2005yq,Leung:2005xz}, we have critically studied a similar problem in (3+1)-dimensional QED in a strong external magnetic field. In particular, we carefully examined the consistency of the bare vertex approximation (or the so-called rainbow approximation) that has been extensively used in truncating the SD equations in the lowest Landau level approximation. Based on the gauge independent analysis~\cite{Leung:2005yq,Leung:2005xz}, we argue that the analysis done in
Ref.~\cite{Bashir:2005wt} is conceptually incorrect and hence the main conclusions thereof are not valid.

Several important lessons regarding the consistent truncation of the SD equations can be learned from the gauge independent analysis presented in Refs.~\cite{Leung:2005yq,Leung:2005xz}. (i) A consistent truncation of the SD equations is the truncation which respects the Ward-Takahashi (WT) identity satisfied by the truncated fermion-boson vertex and inverse fermion propagator. (ii) In a consistent truncation of the SD equations, the WT identity is a \emph{necessary} condition for establishing the gauge independence of physical observables calculated therein, but it is far from sufficient. (iii) The WT identity guarantees that the consistently truncated vacuum polarization is transverse (hence gauge independent), and that the gauge dependent part of the consistently truncated fermion self-energy is proportional to the full inverse fermion propagator (hence vanishing on the fermion mass shell).

In general, the dynamically generated fermion mass and chiral condensate are the two physical observables associated with dynamical chiral symmetry breaking. We first discuss briefly the case of dynamical fermion mass. The interested reader is referred to Refs.~\cite{Leung:2005yq,Leung:2005xz} for further discussions. We have explicitly demonstrated~\cite{Leung:2005yq,Leung:2005xz} that in order to obtain the gauge independent, physical dynamical fermion mass, it is mandatory to evaluate the fermion self-energy \emph{on the fermion mass shell} in a consistent truncation of the SD equations, in which the WT identity is fulfilled. Were the fermion self-energy evaluated \emph{off the fermion mass shell}, even in a consistent truncation of the SD equations the resulting dynamical fermion mass is inevitably be gauge dependent. This is a direct consequence of an important, but largely overlooked, theorem about the systematic expansion (truncation) schemes in gauge theories~\cite{Kobes:1990dc}, which states that in gauge theories the singularity structures (i.e., the positions of poles and branch singularities) of the gauge boson and fermion propagators are gauge independent when all contributions of a given order of a systematic expansion scheme are accounted for. Thus, the gauge independent, physical dynamical fermion mass has to be determined by the pole of the full fermion propagator in a consistent truncation of the SD equations. This is tantamount to evaluating the consistently truncated fermion self-energy on the fermion mass shell.

We now turn to the case of chiral condensate, the physical observable intended to be calculated in Ref.~\cite{Bashir:2005wt}. Because of the off-shell gauge dependence of the fermion self-energy (even in a consistent truncation of the SD equations), the off-shell full fermion propagator is inherently a gauge dependent quantity. Hence the chiral condensate, i.e., the trace of the full fermion propagator, calculated in a certain truncation of the SD equations (be it consistent or not) is inevitably gauge dependent if one simply takes the trace naively and carelessly. Unfortunately, this is what has been done in Ref.~\cite{Bashir:2005wt}. The authors of Ref.~\cite{Bashir:2005wt} were fully aware of the potential gauge dependence of the chiral condensate and they proposed the use of the Landau-Khalatnikov-Fradkin (LKF) transformation as a remedy for the problem. But without addressing the inherent gauge dependence of the full fermion propagator, this is a futile effort. While the LKF transformation preserves the WT identity in different covariant gauges, it does not render the full fermion propagator gauge independent (even in a consistent truncation of the SD equations). This is because, as we have emphasized above, the WT identity is a necessary (but not sufficient) condition for establishing the gauge independence of physical observables in that it guarantees the gauge independence of the on-shell fermion self-energy but not of the off-shell one. Furthermore, since the input to the LKF transformation, i.e., the chiral condensate calculated in the Landau gauge, is not rendered gauge independent in the first place, the LKF transformation simply cannot automatically yield a gauge independent result for the chiral condensate in other covariant gauges.

A close examination of the numerical results obtained in Ref.~\cite{Bashir:2005wt} reveals several fundamental conceptual contradictions. Since the WT identity in the rainbow approximation is not satisfied in ordinary QED, the rainbow approximation cannot be a consistent truncation of the SD equations in ordinary QED. This is a well-known fact which was also noted by the authors of Ref.~\cite{Bashir:2005wt}. As a consequence, the resulting chiral condensate calculated in the rainbow approximation is inevitably gauge dependent. Nevertheless, as can be seen clearly in Fig.~8, when compared with the expected gauge dependent result obtained by using the inconsistently truncated SD equations, the LKF transformation in the rainbow approximation still yields a seemingly ``gauge independent'' chiral condensate. A similar contradiction is also evident in Fig.~9, in which the LKF transformation again gives rise to a seemingly ``gauge independent'' result for the dynamical fermion mass, another physical observable known to be gauge dependent when calculated in the rainbow approximation. These numerical results, together with those for the chiral condensate obtained in other truncations (see Fig.~11), indicate that under the LKF transformation implemented numerically in a certain truncation (be it consistent or not), the \emph{numerical values} of the yet gauge dependent chiral condensate and dynamical fermion mass calculated in the Landau gauge are approximately preserved in an arbitrary covariant gauge. Consequently, Eq.~(10) should be interpreted as the statement that under the LKF transformation the numerical value of the chiral condensate in an arbitrary covariant gauge is exactly the same as that calculated in the Landau gauge rather than the statement, as was done by the authors of Ref.~\cite{Bashir:2005wt}, that the LKF transformation ``ensures the gauge invariance of the chiral condensate'' in an arbitrary covariant gauge. It is precisely this misinterpretation that leads the authors to their incorrect conclusions.

In conclusion, we stress that in gauge theories any analysis of the SD equations should not and cannot be considered reliable or complete unless the gauge independence of physical observables calculated therein is unambiguously demonstrated. Therefore, if the authors of Ref.~\cite{Bashir:2005wt} would like to claim the validity of their conclusions then it remains their responsibility to demonstrate explicitly as well as unequivocally the gauge independence of the results they obtained.

This work was supported in part by the National Science Council of Taiwan under grants 95-2112-M-032-010 and 96-2112-M-032-005-MY3.

\end{document}